From R.R.Khuri@qmw.ac.uk Thu Apr  9 16:03:06 1998
Date: Wed, 8 Apr 1998 18:06:04 +0100 (BST)
From: Ramzi Khuri <R.R.Khuri@qmw.ac.uk>
To: Nemanja Kaloper <kaloper@leland.Stanford.EDU>
Cc: Ramzi Khuri <R.R.Khuri@qmw.ac.uk>, rcm@itp.ucsb.edu
Subject: Re: PLB 12862 (fwd) - accepted

I vote to say what you said "version accepted in PLB, minor changes".
I think Chris Pope would probably prefer it that way, since it won't
make him seem fussy.

Below is the latest version I have, except that for some reason it's
an editted version of the file you originally sent to hepth, and not the
one we would retrieve directly from hepth (in other words I had to insert
by hand the hepth number). That doesn't matter, does it? 

Chow,

Ramzi

\documentstyle[12pt]{article}
\def\be{\begin{equation}}
\def\ee{\end{equation}}
\def\ba{\begin{eqnarray}}
\def\ea{\end{eqnarray}}
\def\bq{\begin{quote}}
\def\eq{\end{quote}}

\newcommand{\labell}[1]{\label{#1}} 
\newcommand{\reef}[1]{(\ref{#1})}

\def\part{\partial}

\def\beq{\begin{equation}}
\def\eeq{\end{equation}}
\def\beqa{\begin{eqnarray}}
\def\eeqa{\end{eqnarray}}

\def\F{{\cal F}}
\def\A{{\cal A}}
\def\tA{\tilde{\cal A}}
\def\ie{{\it i.e.,}\ }
\def\eg{{\it e.g.,}\ }
\def\iden{1\!\!1}

\begin{document}

\thispagestyle{empty}
\rightline{\small hep-th/9803066 \hfill UCSB NSF-ITP-98-023}
\rightline{\small \hfill Stanford SU-ITP-98-08}
\rightline{\small \hfill McGill/98-04}
\rightline{\small \hfill QMW-PH-98-08}
\vspace*{1cm}
\begin{center}
{\Large \bf On Generalized Axion Reductions}\\
\vspace*{1cm}
Nemanja Kaloper\footnote{E-mail:
kaloper@leland.stanford.edu}\\
\vspace*{0.2cm}
{\it Department of Physics, Stanford University}\\
{\it Stanford, CA 94305-4060, USA}\\
\vspace*{0.4cm}
Ramzi R. Khuri\footnote{E-mail: R.R.Khuri@qmw.ac.uk}\\
\vspace*{0.2cm}
{\it Department of Physics, Queen Mary and Westfield College}\\
{\it Mile End Road, London E1 4NS, UK}\\
\vspace*{0.4cm}
Robert C. Myers\footnote{E-mail: rcm@itp.ucsb.edu}$^,$\footnote{Permanent
address: Department of Physics,
McGill University, Montr\'eal, PQ, H3A 2T8, Canada}\\
\vspace*{0.2cm}
{\it Institute of Theoretical Physics,
University of California}\\
{\it Santa Barbara, CA 93106-4030, USA}\\
\vspace{2cm}
ABSTRACT
\end{center}
Recently interest in using generalized reductions to construct
massive supergravity theories has been revived in the context
of M-theory and superstring theory. These compactifications
produce mass parameters by introducing a linear dependence
on internal coordinates in various axionic fields. Here we point out
that by extending the form of this simple ansatz, it is always
possible to introduce the various mass parameters simultaneously.
This suggests that the various ``distinct'' massive
supergravities in the literature should all be a part of
a single massive theory.
\vfill
\setcounter{page}{0}
\setcounter{footnote}{0}
\newpage

Dimensional reduction provides an important window on the
duality relations amongst the various superstring theories,
as well as eleven-dimensional supergravity. Recently 
generalized Scherk-Schwarz reductions\cite{ss}
have received a renewed interest\cite{novel,cow,others,berg,genera}.
This activity began with the remarkable discovery\cite{novel}
that the massive IIa supergravity of Romans\cite{larry} is related
by T-duality to a Scherk-Schwarz compactification of the massless IIb theory.
This result then provides a massive extension of the standard T-duality between
type IIa and IIb superstring theories compactified on $S^1$\cite{bigg}.
Further, renewed interest stems from the recent investigations
of extended objects in string theory. Massive supergravities are
particularly relevant in the case of domain walls\cite{cow,others}.
Some earlier investigations of Scherk-Schwarz reductions in string theory
were made
both at the level of the low-energy supergravity action\cite{low},
and at the level of the world-sheet conformal field theory\cite{ryan}.

The key to the generalized Scherk-Schwarz reductions\cite{ss,genera} is that,
using global symmetries arising in a compactification,
the fields may be given a (specific) dependence on the internal
coordinates. However, the resulting theory is still independent of all
of the internal coordinates. The recent discussions\cite{novel,cow,others,berg}
in the context of low-energy string or M-theory
focus on toroidal compactifications and various axionic symmetries,
\ie constant shifts of certain scalar fields. In the simplest
cases then, the axions appear in the action covered
by derivatives, \ie the scalar field $\chi$ appears everywhere in the action
only as $\partial_\mu\chi$, or in form notation as $d\chi$. If upon
compactification such axions are
given a linear dependence on the internal coordinates, only the slope
of this dependence appears in the reduced action\cite{cow}, \ie
\beq
\chi(x,z)=\chi(x)+m\,z\qquad  \longrightarrow 
\qquad d\chi(x,z)=d\chi(x)+m\,dz
\labell{simple}
\eeq
The slope parameters then play
the role of masses in the compactified theory.

A fundamental
axion scalar appears in the ten-dimensional IIb supergravity and plays
the central role in the T-duality to the massive IIa theory\cite{novel}.
In general, however, the axions of interest arise in
a partially reduced theory as internal components of gauge fields, form-fields
or the metric. The translation symmetry of these scalars is then a residue of
a local gauge invariance in the uncompactified theory. Introducing the
linear ansatz can also then be regarded as giving an expectation to
certain field strengths on the internal space, or introducing a twist
or curvature in the internal geometry. Further,
these axions may have nonderivative couplings through
descendants of `Chern-Simons' interactions in the unreduced theory.
Introducing the linear ansatz \reef{simple} then requires certain
field redefinitions to cover the appropriate scalars with
derivatives\cite{cow}. As a result, however, a conflict may arise in 
simultaneously introducing the linear ansatz for several different
axions. Below, we show that this conflict can be resolved by
the introduction of a slightly
generalized ansatz, which is quadratic (or higher order) in 
the internal coordinates. Field redefinitions
may be found to reduce the internal dependence to a linear one, although
still not of the simple form given in eq. \reef{simple}. 

In this letter, we make the discussion explicit by referring to
a specific example considered in ref.~\cite{cow}. 
Cowdall {\it et al.} \cite{cow} applied Scherk-Schwarz reductions 
to eleven-dimensional
supergravity to produce a variety of maximally-supersymmetric massive
supergravities in $D\le8$. They discussed the case of simultaneously
applying the linear ansatz \reef{simple} to several axions, but were
limited by the problem discussed above. The present discussion provides
an explicit extension of their results, and
generalizing this approach 
to other cases is a straightforward exercise.

In the toroidal compactification of eleven-dimensional supergravity
to $D=8$, three axions $\A_0^{({ij})}$ (with $i,j=1,2,3$ and $i<j$)
appear in the off-diagonal components of the internal metric. The
appropriate dreibein on the internal torus
may be written (using the notation of
\cite{pope}\footnote{We have simplified this notation with respect
to the scalars $\phi_i$, which do not play an important role in the
following.}):
\beq
e^A{}_M=\pmatrix{e^{-\phi_1}&e^{-\phi_1}\A_0^{(12)}&e^{-\phi_1}\A_0^{(13)}\cr
                 0&e^{-\phi_2}&e^{-\phi_2}\A_0^{(23)}\cr
                 0&0&e^{-\phi_3}\cr}
\labell{three}
\eeq
where $A$ and $M$ denote the tangent-space and holonomic
indices, respectively.
The kinetic terms of the axions are governed by
the ``field strengths"
\beq
\F_1^{(12)}=d\A_0^{(12)}
\qquad
\F_1^{(13)}=d\A_0^{(13)}-\A_0^{(23)}d\A_0^{(12)}
\qquad
\F_1^{(23)}=d\A_0^{(23)}
\labell{strong}
\eeq
It is clear here that upon compactifying to $D=7$ one can 
straightforwardly introduce the linear ansatz \reef{simple} for $\A_0^{(12)}$
and $\A_0^{(13)}$.
To apply this ansatz to $\A_0^{(23)}$, one must redefine the
fields\cite{cow} as $\tA_0^{(13)}=\A_0^{(13)}-\A_0^{(23)}\A_0^{(12)}$,
such that
\beq
\F_1^{(13)}
=d\tA_0^{(13)}+\A_0^{(12)}d\A_0^{(23)}\ .
\labell{edefi}
\eeq
Now in reducing to $D=7$, one can apply the ansatz
\beq
\A_0^{(23)}(x,z)=\A_0^{(23)}(x)+m^{(23)}z.
\labell{simpa}
\eeq
However, from eq. \reef{edefi}, one sees that this ansatz may
no longer be applied to $\A_0^{(12)}$.

As an alternative to making the above field redefinition,
one could extend the compactification ansatz slightly as follows:
\beqa
\A_0^{(12)}(x,z)&=&\A_0^{(12)}(x)
\nonumber\\
\A_0^{(23)}(x,z)&=&\A_0^{(23)}(x)+m^{(23)}z
\nonumber\\
\A_0^{(13)}(x,z)&=&\A_0^{(13)}(x)+m^{(23)}z\,\A_0^{(12)}(x)
\labell{simpb}
\eeqa
The additional term added to $\A_0^{(13)}$ is a reflection of the
fact that the axion shift
symmetry of $\A_0^{(23)}$ in the original theory is accompanied by
a compensating shift of $\A_0^{(13)}$ so as to leave $\F^{(13)}_1$
invariant.
We see by replacing \reef{simpb} into the original
expression for the field strengths \reef{strong}
that all of the explicit $z$
dependence cancels. (Alternatively, we note that this
ansatz is identical to the original one in which implicitly
we have reduced the new axion as $\tA_0^{(13)}(x,z)=\tA_0^{(13)}(x)$.)
While the extended ansatz \reef{simpb} does not resolve the problem of
simultaneously introducing two mass parameters, $m^{(23)}$ and $m^{(12)}$,
it does show that explicitly covering the axions with derivatives is
not essential to introducing the mass parameters. 
This was anticipated in ref.~\cite{genera}, where it was noted that the
Scherk-Schwarz construction\cite{ss} applies for general global symmetries.
Thus one might believe
that a modified ansatz would allow for the simultaneous inclusion
of both parameters. While we originally constructed such an extended
ansatz by trial and error, in fact, it appears quite naturally using
the full formalism originally developed by Scherk and Schwarz\cite{ss}.

Within the Scherk-Schwarz formalism, one begins by identifying
the relevant global symmetries.
Here, they are a part of the $SL(3,R)$ symmetry acting on the internal
three-torus, which acts on the dreibein
\reef{three} as $e^A{}_M\rightarrow e^A{}_N\,T^N{}_M$.
The translations of the axions can be identified as the 
three transformations with generators
\beq
M^{(12)}=\pmatrix{0&1&0\cr 0&0&0\cr 0&0&0\cr}
\qquad
M^{(13)}=\pmatrix{0&0&1\cr 0&0&0\cr 0&0&0\cr}
\qquad
M^{(23)}=\pmatrix{0&0&0\cr 0&0&1\cr 0&0&0\cr}\ .
\labell{generate}
\eeq
For example, $e^A{}_M\rightarrow e^A{}_N\,\exp[\lambda M^{(12)}]^N{}_M$
accomplishes a shift $\A^{(12)}_0\rightarrow\A^{(12)}_0+\lambda$.
Note that $M^{(23)}$ produces $\A^{(23)}_0\rightarrow\A^{(23)}_0+\lambda$,
and as well, the compensating shift
$\A^{(13)}_0\rightarrow\A^{(13)}_0+\lambda\A^{(12)}_0$.
A distinguishing property of these three generators \reef{generate}
is that they are nilpotent.

Now in the Scherk-Schwarz reduction\cite{ss} to $D=7$, 
one introduces the following specific dependence on the new internal
coordinate $z$ into the dreibein \reef{three}:
\beq
e^A{}_N(x,z)=e^A{}_N(x)\,U(z)^N{}_M=e^A{}_N(x)\,\exp[M z ]^N{}_M
\labell{anss}
\eeq
where $M=\sum m^{(ij)}M^{(ij)}$.
If we consider only a single nonvanishing mass parameter at a
time, it is clear that this ansatz reproduces the usual linear ansatz
discussed above because the individual generators are
nilpotent, \ie the exponential reduces to $\iden+M\,z$.
However, in the case that $m^{(12)}$ and $m^{(23)}$ are simultaneously
chosen to be nonvanishing, $(M)^2=m^{(12)}m^{(23)}\,M^{(13)}\ne0$
while $(M)^3=0$. Thus in this situation the linear ansatz is naturally
extended to one quadratic in the internal coordinate $z$.
Explicitly the axions are chosen as:
\beqa
\A_0^{(12)}(x,z)&=&\A_0^{(12)}(x)+m^{(12)}z
\labell{zerof}\\
\A_0^{(23)}(x,z)&=&\A_0^{(23)}(x)+m^{(23)}z
\nonumber\\
\A_0^{(13)}(x,z)&=&\A_0^{(13)}(x)+m^{(13)}z+m^{(23)}z\,\A_0^{(12)}(x)
+{1\over2}m^{(12)}m^{(23)}z^2
\nonumber
\eeqa
One can verify that there is no explicit $z$ dependence in
$\F_1^{(13)}$ with this ansatz.
Thus within the full Scherk-Schwarz framework\cite{ss},
one finds that there is no obstacle to turning on all
of the mass parameters simultaneously. 

These axions also couple to other fields in the $D=8$ supergravity,
and one must also choose a consistent compactification ansatz to
ensure that the corresponding field strengths do not introduce
a $z$ dependence in the compactified theory. The
Scherk-Schwarz formalism provides a precise prescription
to accomplish this result. Essentially any of the fields carrying
internal holonomic indices are also contracted with the same matrix $U$
appearing in eq.~\reef{anss}. In the present case of the
compactification of eleven-dimensional supergravity, one must consider the
components of the three-form potential,
\eg $A_{mN_1N_2}(x)\,U(z)^{N_1}{}_{M_1}U(z)^{N_2}{}_{M_2}$ and
$A_{m_1m_2N}(x)\,U(z)^N{}_M$. Following the notation of \cite{pope},
these correspond to the one-forms $A^{(ij)}_1$ and two-forms
$A^{(i)}_2$. In the end, one arrives at the following reduction
ansatz for the one-forms:
\beqa
A^{(12)}_1(x,z)&=&\pmatrix{A^{(12)}_1(x)\cr A^{(124)}_0(x)\cr}
\qquad\quad
A^{(13)}_1(x,z)=\pmatrix{A^{(13)}_1(x)+m^{(23)}z\,A^{(12)}_1(x)\cr
                           A^{(134)}_0(x)+m^{(23)}z\,A^{(124)}_0(x)\cr}
\labell{onef}\\
A_1^{(23)}(x,z)&=&\pmatrix{A_1^{(23)}(x)+m^{(12)}z\,A_1^{(13)}(x)
-\left(m^{(13)}z-{1\over2}m^{(12)}m^{(23)}z^2\right)A_1^{(12)}(x)\cr
A_0^{(234)}(x)+m^{(12)}z\,A_0^{(134)}(x)
-\left(m^{(13)}z-{1\over2}m^{(12)}m^{(23)}z^2\right)A_0^{(124)}(x)\cr}
\nonumber
\eeqa
and for the two-forms:
\beqa
A_2^{(1)}(x,z)&=&\pmatrix{A_2^{(1)}(x)\cr A_1^{(14)}(x)\cr}
\qquad\quad
A_2^{(2)}(x,z)=\pmatrix{A_2^{(2)}(x)+m^{(12)}z\,A_2^{(1)}(x)\cr
                          A_1^{(24)}(x)+m^{(12)}z\,A_1^{(14)}(x)\cr}
\labell{twof}\\
A_2^{(3)}(x,z)&=&\pmatrix{A_2^{(3)}(x)+m^{(23)}z\,A_2^{(2)}(x)
+\left(m^{(13)}z+{1\over2}m^{(12)}m^{(23)}z^2\right)A_2^{(1)}(x)\cr
A_1^{(34)}(x)+m^{(23)}z\,A_1^{(24)}(x)
+\left(m^{(13)}z+{1\over2}m^{(12)}m^{(23)}z^2\right)A_1^{(14)}(x)\cr}
\nonumber
\eeqa
Here, we see that the quadratic terms make their presence felt
in $A_1^{(23)}$ and $A_2^{(3)}$. Again
one may explicitly verify that with this ansatz no
dependence on $z$ appears in the corresponding field strengths\footnote{These
field strengths are explicitly listed in ref.~\cite{cow}. Note
that it is important to explicitly retain certain higher order terms, \ie
$F^{(3)}_3=dA_2^{(3)}-(\A_0^{(13)}-\A_0^{(12)}\A_0^{(23)})dA_2^{(1)}-
\A_0^{(23)}dA_2^{(2)}+\ldots$.}.
One must also consider the axion $A^{(123)}_0$ which corresponds
to the three-form potential component with three internal indices.
Following the Scherk-Schwarz prescription, the compactification ansatz is
\beq
A_{N_1N_2N_3}(x)\,U(z)^{N_1}{}_{M_1}U(z)^{N_2}{}_{M_2}U(z)^{N_3}{}_{M_3}
=A_{M_1M_2M_3}(x)\,{\rm det} U\ .
\labell{none}
\eeq
However, ${\rm det} U=1$, so this scalar is
unaffected by the above Scherk-Schwarz ansatz. 
In more general settings, one could not expect such
a cancellation to occur. Further, one might also
consider the spacetime vectors arising from the off-diagonal components
of the eleven-dimensional metric. However, with the present notation
of ref. \cite{pope}, one does not introduce any $z$ dependence for these
vectors --- note that the notations of refs.
\cite{pope} and \cite{ss} differ for these fields. 

It should be clear at this point that if we were to extend this discussion
to generalized Scherk-Schwarz compactifications to lower dimensions,
the linear axion ansatz would again be extended to include cubic
and higher order terms in the internal coordinate. We also note, however,
that there does remain the possibility of using field redefinitions
to simplify the ansatz to one
with only linear dependence on the internal coordinates. In the present
example, redefining $\tA_0^{(13)}=\A_0^{(13)}-{1\over2}\A_0^{(12)}\A_0^{(23)}$
eliminates the quadratic terms in the compactification ansatz \reef{zerof}
leaving
\beq
\tA_0^{(13)}(x,z)=\tA_0^{(13)}(x)+m^{(13)}z+{1\over2}m^{(23)}z\A_0^{(12)}(x)
-{1\over2}m^{(12)}z\A_0^{(23)}(x)\ .
\labell{mod}
\eeq
Similarly redefining
\beqa
\tilde{A}_1^{(23)}&=&A_1^{(23)}-{1\over2}\A_0^{(12)}\A_0^{(23)}\,A_1^{(12)}
\nonumber\\
\tilde{A}_2^{(3)}&=&A_2^{(3)}-{1\over2}\A_0^{(12)}\A_0^{(23)}\,A_2^{(1)}
\labell{redef}
\eeqa
removes the quadratic terms from 
eqs.~\reef{onef} and \reef{twof}. Although linear in $z$, this reduction
ansatz still does not take the original simple form of eq.~\reef{simple}.

In summary, one finds that there is no 
obstacle to simultaneously applying a Scherk-Schwarz
reduction for all four of the eight-dimensional axions,
$\A_0^{(12)}$, $\A_0^{(13)}$, $\A_0^{(23)}$ and
$A_0^{(123)}$ --- we have not considered the latter above, but there
is no conflict in introducing $m^{(123)}$ along with any
of the other mass parameters\cite{cow}. The conclusion 
also applies to other compactifications.
Hence, the various massive supergravities
presented in ref.~\cite{cow} as distinct theories
should actually be regarded as belonging to a single family of theories.
After suitable field redefinitions one finds in the present example
that generically there is a
three-parameter scalar potential involving $m^{(123)}$, $m^{(12)}$
and $m^{(23)}$,
while $m^{(13)}$ can be completely removed from the action.
The latter is essentially accomplished by absorbing
$m^{(13)}$ in the expectation value
of the axion $\A_0^{(23)}$ (as long as $m^{(12)}$ is nonvanishing)
\cite{cow}.
Given the Scherk-Schwarz framework, one should be able to extend
this theory further by beginning with the massive type IIa theory
in ten dimensions and compactifying down to seven dimensions. This
would introduce a fourth mass parameter. On the IIb side, this would
correspond to a compactification of the ten-dimensional theory
on $T^3$ which simultaneously introduces
a twist in the RR axion along with a twist in the torus geometry,
as well as constant internal
expectation values of the NS-NS and RR three-form field strengths.
This is likely to be the most general massive seven-dimensional
supergravity which can be produced using the axionic translation
symmetries.

Many aspects of these results apply universally for generalized
axionic reductions. Individually, the axionic symmetries will
correspond to nilpotent generators of the global symmetry group.
Hence the Scherk-Schwarz reduction will coincide with
the simple linear ansatz \reef{simple} when an individual mass
parameter is introduced. However, when several masses are simultaneously
turned on, the reduction ansatz may involve quadratic and higher
order terms as in eq.~\reef{zerof}. These terms result from  the
failure of the various nilpotent generators to commute with each other.

It would be interesting to investigate the interplay of U-duality
with these Scherk-Schwarz reductions --- some
aspects of this issue have been addressed recently, in \cite{ovrut}.
Introducing the mass
parameters generically breaks some part of the global symmetry group
which would otherwise appear in the compactified theory. However,
it should be possible to write the massive theory in a U-duality
invariant form, as long as the symmetry breaking parameters, \ie
the masses, are endowed with the appropriate transformation
properties\cite{us}. Thus, as is standard in spontaneous symmetry
breaking, a broken symmetry will act as a transformation between 
distinct massive theories, or distinct ``vacua'' of the higher dimensional
theory. In the present case, the full supergravity
duality group in seven-dimensions is $SL(5,R)$. While we have argued
that the mass parameters should form a representation of this group,
we have only identified four such parameters for the seven-dimensional
theory. Thus, the full massive theory must contain new masses beyond
those considered here. In the context of the Scherk-Schwarz framework,
it may be that the latter are associated with symmetries other than
the axionic ones identified here, \eg eq. \reef{generate}.
Thus one probably has to extend the reduction ansatz to include
more general global symmetries\cite{ss,genera}
to produce a U-duality invariant form.
Another aspect of these constructions which would be interesting to
study in the context of U-duality is the non-Abelian gauge symmetries
which arise in the Scherk-Schwarz reductions\cite{ss,berg}.

\vspace{1cm}
{\bf Acknowledgements}

N.K. was supported by NSF Grant PHY-9219345. R.R.K. was supported by a PPARC 
Advanced Fellowship and would like to thank the ITP
and the Department of Physics at UCSB
for their hospitality and where part of this work was completed
under support by NSF Grant PHY94-07194. 
R.C.M. was supported by NSERC of Canada and
by NSF Grant PHY94-07194.

\end{document}